\documentclass[%
 reprint,
amsmath,amssymb,
prl,
]{revtex4-1}
\usepackage{graphicx}
\usepackage{dcolumn}
\usepackage{bm}
\usepackage{epstopdf}

\begin{document}

\preprint{APS/123-QED}

\title{Gap structure of FeSe determined by field-angle-resolved specific heat measurements }

\author{Yue Sun,$^1$}
\email{sunyue@issp.u-tokyo.ac.jp}
\author{Shunichiro Kittaka,$^1$ Shota Nakamura,$^1$ Toshiro Sakakibara,$^1$ Koki Irie,$^2$ Takuya Nomoto,$^3$ Kazushige Machida,$^2$ Jingting Chen,$^4$ and Tsuyoshi Tamegai$^4$}

\affiliation{%
$^1$Institute for Solid State Physics (ISSP), The University of Tokyo, Kashiwa, Chiba 277-8581, Japan\\
$^2$Department of Physics, Ritsumeikan University, Kusatsu, Shiga 525-8577, Japan\\
$^3$RIKEN Center for Emergent Matter Science (CEMS), Hirosawa, Wako, Saitama 351-0198, Japan\\
$^4$Department of Applied Physics, The University of Tokyo, Bunkyo-ku, Tokyo 113-8656, Japan}


\begin{abstract}
Quasiparticle excitations in FeSe were studied by means of specific heat ($C$) measurements on a high-quality single crystal under rotating magnetic fields. The field dependence of $C$ shows three-stage behavior with different slopes, indicating the existence of three gaps ($\Delta_1$, $\Delta_2$, and $\Delta_3$). In the low-temperature and low-field region, the azimuthal-angle ($\phi$) dependence of $C$ shows a four-fold symmetric oscillation with sign change. On the other hand, the polar-angle ($\theta$) dependence manifests as an anisotropy-inverted two-fold symmetry with unusual shoulder behavior. Combining the angle-resolved results and the theoretical calculation, the smaller gap $\Delta_1$ is proved to have two vertical-line nodes or gap minima along the $k_z$ direction, and is determined to reside on the electron-type $\varepsilon$ band. $\Delta_2$ is found to be related to the electron-type $\delta$ band, and is isotropic in the $ab$-plane but largely anisotropic out of the plane. $\Delta_3$ residing on the hole-type $\alpha$ band shows a small out-of-plane anisotropy with a strong Pauli-paramagnetic effect.

\end{abstract}

\maketitle
Superconducting (SC) gap structures are intimately related to the pairing mechanism, which is pivotal for high-temperature superconductors. This issue is crucial for FeSe because of the unexpectedly high superconducting temperature, $T_{\rm c}$, in this system. Although the initial $T_{\rm c}$ is below 10 K \cite{HsuFongChiFeSediscovery}, it can be easily enhanced to 37 K under pressure, \cite{MedvedevNatMat,MargadonnaPRB} and to over 40 K by intercalating spacer layers \cite{BurrardNatMat}. Recently, a monolayer of FeSe grown on SrTiO$_3$ has even shown a sign of $T_{\rm c}$ over 100 K \cite{WangCPLMonolayerFeSe,GeNatMatter}. FeSe manifests some intriguing properties, including a nematic state without long-range magnetic order \cite{McQueenPRL}, crossover from Bardeen-Cooper-Schrieffer (BCS) to Bose-Einstein-condensation (BEC) \cite{Kasahara18112014}, and a Dirac-cone-like state \cite{ZhangFeSeDirac,KontariDiracconePhysRevLett,SunPhysRevB.93.104502}, which are crucial to understanding high-$T_{\rm c}$ superconductivity.

Efforts have been made to elucidate FeSe's gap structure, and the presence of nodes \cite{SongScience,Kasahara18112014} or deep minima \cite{LinFeSeSHPRB,LinJiaoSciRep,hopePhysRevLett,AbdelHcFeSePRB} have been proposed. Even if the multi-gap structure is established \cite{TerashimaPRB,WatsonPRB2016}, gap nodes or minima must still be located in the corresponding bands. Unfortunately, there have been few reports on this issue except for the recent Bogoliubov quasiparticle interference (BQPI) experiments, which found gap minima in both the $\alpha$ and $\varepsilon$ bands \cite{BPQIarxiv}. However, there is still no bulk evidence of the locations of nodes or gap minima, and also no information about the gap from the $\delta$ band. More importantly, details of gap structure, including the three-dimensional (3D) locations of  gap nodes or minima, remain unexplored. To solve these issues, a bulk technique capable of probing quasiparticle (QP) excitations with 3D angular resolution is needed. Field-angle-resolved specific heat (ARSH) measurement is an ideal tool for probing the density of states (DOS) of QPs without interference from surface effects; it is angle resolved because the low-lying QP excitations near the gap nodes (minima) are field-orientation dependent \cite{SakakibaraReview}. ARSH measurements have been well applied to investigating the locations and types of nodes (gap minima) in heavy-fermion \cite{SakakibaraReview}, spin-triplet \cite{SrRuOASHPRL}, and topological superconductors \cite{YonezawaNatPhy}.

In this Rapid Communication, we investigate the gap structure of FeSe by ARSH measurements. The presence of three gaps is confirmed. The smallest gap, $\Delta_1$, is found to reside on the electron-type $\varepsilon$ band, with two vertical-line nodes (gap minima) along the $k_z$-direction. The intermediate $\Delta_2$ is found to be related to the electron-type $\delta$ band, and is isotropic in the $ab$-plane, but largely anisotropic out-of-plane. The largest $\Delta_3$ residing in the hole-type $\alpha$ band shows a small out-of-plane anisotropy with a strong Pauli-paramagnetic effect.

High-quality FeSe single crystals were grown by the vapor transport method \cite{SunPhysRevBJcFeSe,SunPhysRevB.93.104502}. The temperature dependence of the specific heat was measured using a physical-property-measurement system (PPMS) under fields up to 9 T. The magnetic-field dependence of the specific heat was measured in a dilution refrigerator under fields up to 14.7 T. The field-orientation dependence of the specific heat was measured in an 8 T split-pair superconducting magnet with a $^3$He refrigerator. The refrigerator can be continuously rotated by a motor on top of the Dewar with an angular resolution better than 0.01$^\circ$. The directions of the crystal axes were determined by single-crystal X-ray diffraction. The definitions of the in-plane (azimuthal) angle $\phi$ and the out-of-plane (polar) angle $\theta$ are shown in Fig. 1(a). Due to the presence of twin boundaries (TBs), the $a$- and $b$-axes cannot be distinguished. The TBs' effects will be discussed later.

\begin{figure}\center
\includegraphics[width=8cm]{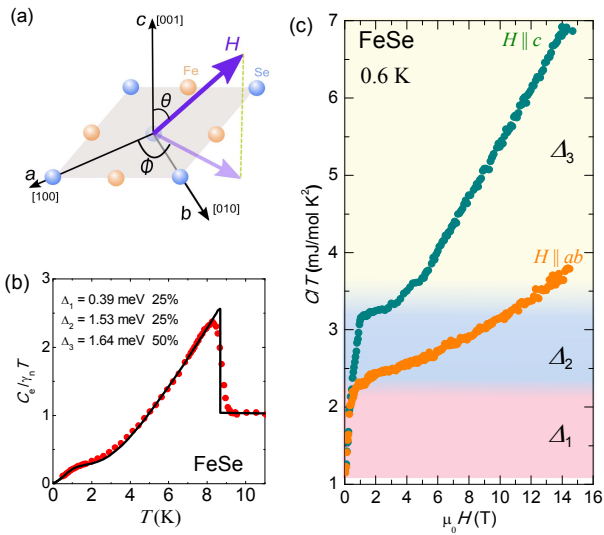}\\
\caption{(a) Definitions of azimuthal ($\phi$) and polar ($\theta$) angles with respect to the crystal axes. The Se in the center is on the layer below. (b) Normalized zero-field electronic-specific heat, $C_e/\gamma_nT$ vs $T$, together with the fit by a three-gap model. (c) Magnetic-field dependences of $C$($H$)/$T$ for $H\parallel  c$ and $H\parallel  ab$ at 0.6 K.}\label{}
\end{figure}

The obtained single crystal shows very high quality with a sharp SC-transition width, an extremely small residual resistivity $\rho_0$ $\sim$ 1 $\mu\Omega$ cm, and a large residual-resistivity ratio (RRR) over 400, as shown in Supplementary S1 \cite{supplement}. The zero-field electronic specific heat $C_{\rm e}/T$ obtained by subtracting the phonon terms (Supplementary S2 \cite{supplement}), is shown in Fig. 1(b). Besides the specific-heat jump at $T_c$, a hump-like behavior below 2 K is also observed, which is typical of multi-gap superconductors like MgB$_2$ and Lu$_2$Fe$_3$Si$_5$ \cite{MgB2PhysRevLetttwogap,*NakajimaRuFeSitwogapPhysRevLett.100.157001}. The fitting of $C_{\rm e}$ will be discussed later.

\begin{figure*}\center
\includegraphics[width=15cm]{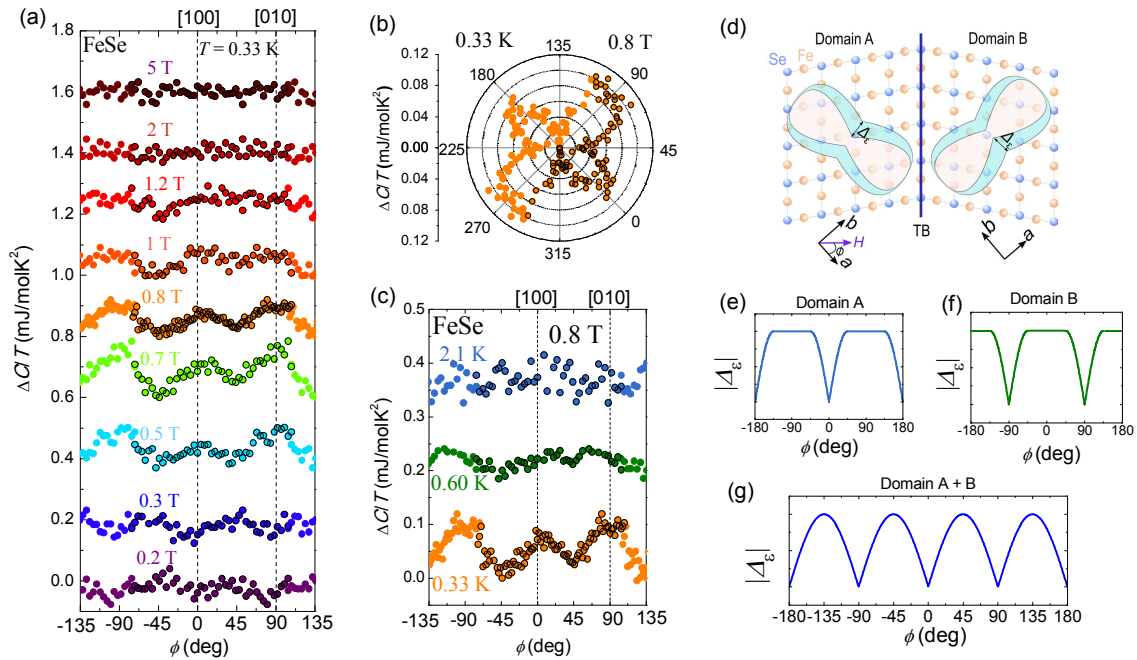}\\
\caption{(a) Azimuthal angle dependence of the specific heat $\Delta C(\phi)/T$ measured under various fields at 0.33 K. $\Delta C(\phi)/T$ is defined as $C(\phi)/T$-$C(-45^\circ)/T$, and each subsequent curve is shifted vertically by 0.2 mJ/molK$^2$. Symbols with black outlines are measured data, and those without are mirrored points to show the symmetry. (b) Polar plot of the $\Delta C(\phi)/T$ at 0.33 K under 0.8 T. (c) $\Delta C(\phi)/T$ measured under 0.8 T at 0.33 K, 0.60 K, and 2.1 K. Each subsequent curve is shifted by 0.2 mJ/molK$^2$. (d) In-plane schematic view of the atomic arrangement and gap structure of $\Delta_\varepsilon$ \cite{BPQIarxiv} in domains A and B sandwiching a TB. Schematic-gap functions of $\Delta_\varepsilon$ in domains (e) A and (f) B, and (g) their superpositions.}\label{}
\end{figure*}

For a superconductor with an isotropic single gap, $C$($H$)/$T$ is linearly related to $H$ because the low-energy QPs are mainly localized in the vortex core, whose density is proportional to $H$ \cite{NakaiPhysRevB.70.100503}. For a single gap with nodes, $C$($H$)/$T$ is usually proportional to $H^{1/2}$ because of the QPs' Doppler shift caused by supercurrents around the vortex core \cite{VolovikJETPLett}. In between, $C$($H$)/$T$ in an anisotropic gap shows a crossover from linear-$H$ to $H^{1/2}$ dependence \cite{NakaiPhysRevB.70.100503}.

The $C$($H$)/$T$ values of FeSe at 0.6 K for $H \parallel c$ and $H \parallel  ab$ measured up to 14.7 T are shown in Fig. 1(c). Their behavior differs from the situations discussed above for a single gap. $C$($H$)/$T$ manifests different slopes depending upon the magnitude and direction of the field. Below $\sim$2.2 mJ/molK$^2$, $C$($H$)/$T$ increases rapidly with the same slope for both $H\parallel c$ and $H\parallel ab$. Then, $C(H\parallel c)/T$ continues to increase with a slightly reduced slope up to $\sim$3.2 mJ/molK$^2$, while the slope of $C$($H\parallel ab$)/$T$ is largely suppressed. As $H$ further increases, $C$($H\parallel c$)/$T$ increases to $H_{\rm c2}$ ($\sim$14 T) with a sudden reduced smaller slope, while the slope for $C$($H\parallel ab$)/$T$ is gradually enhanced. Since the value of $C$($H$)/$T$ is proportional to the DOS of the QPs, the sudden suppression of the slope usually indicates closing of a gap, as discussed in MgB$_2$ \cite{MgB2PhysRevLetttwogap}. In FeSe, $C$($H$)/$T$ shows obvious three-stage behavior with different slopes, which implies that FeSe contains three gaps labeled $\Delta_1$, $\Delta_2$, and $\Delta_3$. Their dominant regions are indicated by different colors. The three-gap structure is consistent with its band structure \cite{TerashimaPRB,WatsonPRB2016}.

The $C$($H$)/$T$ for two field orientations also carries information concerning the out-of-plane anisotropy. $\Delta_1$ is nearly isotropic, since $C$($H\parallel c$)/$T$ and $C$($H\parallel ab$)/$T$ have similar slopes. By contrast, $\Delta_2$ is obviously strongly anisotropic. For $\Delta_3$, we cannot draw a solid conclusion from the present data, since $H_{c2}^{ab}$ is much larger than our experimental limit. However, considering that is $\sim$30 T from Ref. \cite{TerashimaPRB}, we estimate the out-of-plane anisotropy for $\Delta_3$ to be $H_{c2}^{ab}$/$H_{c2}^{c}$ $\sim$2.

ARSH is used to study the wave-vector-dependent gap structure of FeSe in more detail. Figure 2(a) shows the azimuthal angle-resolved $\Delta C(\phi)/T$ at 0.33 K. $\Delta C(\phi)/T$ manifests a four-fold symmetry under small fields, more easily recognized in the polar plot (Fig. 2(b)). Below 0.5 T, $\Delta C(\phi)/T$ shows minima for the $H\|[100]$ and [010] axes ($\phi=0^\circ$ and $90^\circ$) and maxima for the $H\|[110]$ axis ($\phi=45^\circ$). At $H \geq $ 0.5 T, $\Delta C(\phi)/T$ becomes maximal for the $H\|[100]$ and [010] axes, but minimal for the $H\|[110]$ axis. The isotropic $s$-wave does not cause this behavior, as it should be angularly independent.

In a nodal gap under small fields, the zero-energy DOS in the vortex state is mainly induced by the Doppler shift, $\delta E = m_e \textbf{\emph{v}}_F \cdot \textbf{\emph{v}}_s$, of the excited QPs around nodes, where $m_e$ is the electron mass,  $\emph{\textbf{v}}_F$ the Fermi velocity, and $\emph{\textbf{v}}_s$ the local superfluid velocity perpendicular to the field \cite{VolovikJETPLett}. Thus, the Doppler shift energy, $\delta E$, depends upon the direction of the field with respect to the nodal position. When the field is parallel to $\emph{\textbf{v}}_F$ around the node ($H\parallel$ node), the zero-energy DOS is small because $\delta E=0$ at the node. By contrast, it will be largely enhanced when the field is perpendicular to $\emph{\textbf{v}}_F$ ($H\perp$ node) because $\delta E$ becomes maximal. Therefore, in the low-field region, specific heat shows minima for $H\parallel$ node, and maxima for $H\perp$ node. On the other hand, under high fields, the QP scattering by the magnetic field is strongly enhanced. In this case, a much higher finite-energy DOS around the nodal positions will be excited for $H\parallel$ node. When the finite-energy DOS overcomes the zero-energy DOS, the oscillation switches signs, i.e., specific heat becomes maxima for $H\parallel$ node, but minima for $H\perp$ node \cite{VorontsovPhysRevLett.96.237001,HiragiJPSJASHcal}. Such sign change is common for nodal superconductors such as CeCoIn$_5$ \cite{AnPhysRevLett.104.037002} and KFe$_2$As$_2$ \cite{KittakaKFe2As2JPSJ}. In FeSe, this behavior is also observed when the field crosses $\sim$0.5 T (Fig. 2(a)). We cannot distinguish nodes from gap minima since the anisotropic $s$-wave with gap minima also manifests similar oscillatory behavior \cite{KittakaCeRu2JPSJ.82.123706}. Based on the discussion above, the nodes or gap minima are found to reside in the $<$100$>$ and/or $<$010$>$ directions. Furthermore, oscillations with sign change are only observed under small fields at low temperatures. As shown in Figs. 2(a) and 2(c), the four-fold symmetry is almost smeared when the field exceeds 1 T or temperature reaches $\sim$2 K, suggesting the observed nodes (gap minima) are from the smaller gap $\Delta_1$.

Before discussing the topology and origin of gap nodes (minima) in $\Delta_1$, we first point out the effects of the TBs, formed when the crystal transitions from tetragonal to orthorhombic structure. In neighboring regions across the TB, the  in-plane lattice is rotated by 90$^\circ$, as shown schematically in Fig. 2(d) and directly observed by scanning tunneling microscopy \cite{WatashigePRX}. TBs are almost unavoidable and exist in our crystals (Supplementary S3 \cite{supplement}).

\begin{figure*}\center
\includegraphics[width=15cm]{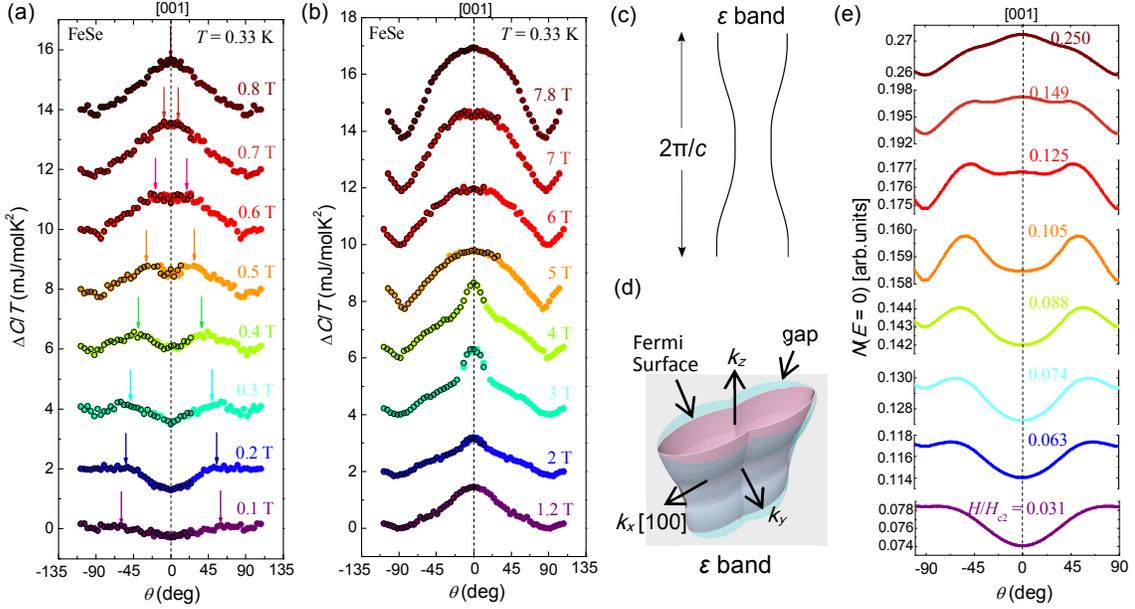}\\
\caption{Polar angle dependence of the specific heat $\Delta C(\theta)/T$ measured under fields (a) below and (b) above 1 T at 0.33 K. $\Delta C(\theta)/T$ is defined as $C(\theta)/T$-$C(-90^\circ)/T$, and each subsequent curve is shifted vertically by 2 mJ/molK$^2$. Black-outlined symbols are the measured data; the others are mirrored points to show the symmetry more clearly. (c) Calculated Fermi surface cross section for the $\varepsilon$ band along the $k_z$-direction. (d) Schematic-gap structure for the $\varepsilon$ band in 3D. (e) Calculated results of the $\theta$-dependence of the zero-energy DOS, $N(E=0)$, in magnetic fields normalized by $H_{c2}$.}\label{}
\end{figure*}

FeSe is experimentally determined to consist of one hole-type $\alpha$ band located at the $\Gamma$ point and two electron-type bands, $\delta$ and $\varepsilon$, at the M point \cite{TerashimaPRB,WatsonPRB2016}. Although four-fold symmetry is observed, the $d$-wave-like gap structure seems unlikely because it is inconsistent with most previous reports \cite{BPQIarxiv,XuAPREFeSeSPhysRevLett.117.157003}. Recent BQPI measurements found that the gaps from the $\varepsilon$($\Delta_\varepsilon$) and $\alpha$($\Delta_\alpha$) bands are two-fold symmetric but with different gap functions \cite{BPQIarxiv}. The schematic form of $\Delta_\varepsilon$ is shown in Fig. 2(d). Considering the TBs, $\Delta_\varepsilon$ in the neighboring domains A and B is  rotated by 90$^\circ$ and the corresponding gap functions are also shifted by 90$^\circ$ as shown in Figs. 2(e) and 2(f), respectively. Superposition of the two gap functions is shown in Fig. 2(g) by simply assuming an equal ratio, which shows four-fold symmetry similar to the experimental observation of $\Delta C(\phi)/T$. On the other hand, $\Delta_\alpha$ manifests a waveform of $\cos2\phi$ \cite{BPQIarxiv}, whose oscillation differs from the four-fold symmetry, when the effects of TBs are considered (Supplementary S4 \cite{supplement}). Hence, the $\Delta_1$ with nodes (gap minima) is assigned to the electron-type $\varepsilon$ band.  The four-fold symmetric ARSH has also been observed in FeTe$_{0.55}$Se$_{0.45}$ \cite{Zengnacomm}. However, it originates from a large anisotropic gap with four minima, which is different from the present case in FeSe.

To determine the 3D gap structure of FeSe, we performed out-of-plane ARSH measurements. $\Delta C(\theta)/T$ ($\phi$ = 45$^\circ$) at 0.33 K is shown in Fig. 3(a) - (b) under various fields. In the $\mu_0H<$ 1 T ($\Delta_1$ dominant) region, $\Delta C(\theta)/T$ first shows minima in the $<$001$>$ direction ($H\parallel c$) with two shoulders under small fields, as indicated by arrows. With increasing field, the minima gradually increase, and the two shoulders move toward the $<$001$>$ direction. Finally, the minima at $<$001$>$ turn to maxima and the two shoulders converge into one maximum at $\sim$0.8 T.

To understand the anisotropy-inversion behavior of $\Delta C(\theta)/T$, we performed microscopic calculations of the DOS using quasi-classical Eilenberger theory within the Kramer-Pesch approximation \cite{TsutsumipolarCcalculationPhysRevB.94.224503}. To simulate the cylindrical open Fermi surface of the $\varepsilon$ band observed in Ref. \cite{TerashimaPRB} (consisting of parallel and warped segments, as shown in Fig. 3(c)), we model the Fermi velocity, $v_z$($k_z$), along $k_z$ as $v_z$($k_z$)$\propto\sin^5ck_z$ ($-\pi/c\leq k_z\leq\pi/c$ with $c$ being the lattice constant along $k_z$), which has a substantial parallel segment. Assuming two vertical-line nodes (gap minima) along $k_z$, the calculation result is presented in Fig. 3(e). It is qualitatively similar to the experimental results. The anisotropy-inverted $\Delta C(\theta)/T$ can also be explained by competition between the zero-energy and finite-energy DOSs based on the Doppler shift effect. In the case of vertical-line nodes (gap minima), $\textbf{\emph{v}}_F^{H\parallel c}\cdot\textbf{\emph{v}}_s^{H\parallel c}<\textbf{\emph{v}}_F^{H\parallel ab}\cdot\textbf{\emph{v}}_s^{H\parallel ab}$ in the small-field region because of $k_z$-direction Fermi-surface warping. Under higher fields, the scattering of QPs is largely enhanced for $H\parallel$ nodal (gap minima) lines, making $\textbf{\emph{v}}_F^{H\parallel c}\cdot\textbf{\emph{v}}_s^{H\parallel c}>\textbf{\emph{v}}_F^{H\parallel ab}\cdot\textbf{\emph{v}}_s^{H\parallel ab}$. By contrast, the anisotropy-inversion behavior should be opposite in the case of in-plane point nodes (gap minima). In this case, the Doppler-shift energy, $\delta E$, is maximal for $H\parallel c$ under small fields because $H$ is always perpendicular to $\textbf{\emph{v}}_F$. Contrarily, it is minimal when $H\parallel ab$ because a small angle exists between $H$ and $\textbf{\emph{v}}_F$ \cite{TsutsumipolarCcalculationPhysRevB.94.224503}. Based on the discussion above, we conclude that the $\Delta_1$ contains two vertical-line nodes (gap minima) along the $k_z$-direction, indicated by the 3D schematic-gap structure in Fig. 3(d).

For $\mu_0H>$ 1 T, $\Delta C(\theta)/T$ shows two-fold symmetry in general, reflecting the out-of-plane anisotropy. However, from 2 to 4 T, small peaks in the $<$001$>$ direction can be observed. These may originate from the strong Pauli-paramagnetic effect \cite{TsutsumiPaulilimitPhysRevB.92.020502}. In multi-gap superconductors, moderate magnetic fields enhance the finite-energy QPs as observed in CeCu$_2$Si$_2$ \cite{KittakaCeCu2Si2PhysRevLett.112.067002} and KFe$_2$As$_2$ \cite{KittakaKFe2As2JPSJ}. Since such behavior is only observed for $H\parallel c$ in middle-field range, it is attributed to $\Delta_3$.

Since neither $\Delta C(\phi)/T$ nor $\Delta C(\theta)/T$ in the $\Delta_2$-dominant region (1 T$<\mu_0H<$ 10 T for $H\parallel ab$) show in-plane-anisotropy-related oscillation, $\Delta_2$ is isotropic in the $ab$-plane, which differs from the largely in-plane anisotropic $\Delta_\alpha$ observed in BQPI measurements \cite{BPQIarxiv}. Hence, we conclude that $\Delta_2$ is related to the electron-type $\delta$ band. Thus, the remaining $\Delta_3$ should correspond to the hole-type $\Delta_\alpha$. Since the field in the $\Delta_3$-dominant region is beyond the upper limit of our ARSH measurement system, we lack direct evidence for the topology of  $\Delta_3$. It may be in-plane isotropic or anisotropic with two-fold symmetry based on BQPI results \cite{BPQIarxiv}.

Finally, we return to the temperature dependence of $C_e/T$ shown in Fig. 1(b). Based on the above discussions, we fit the $C_e/T$ with a three-gap model based on the BCS theory by simply assuming $\Delta_1$ has two line nodes; $\Delta_2$ and $\Delta_3$ are both isotropic $s$-waves. In this case, $C_e$=$\gamma_1C_1(\Delta_1\cos2\phi)$+$\gamma_2 C_2(\Delta_2)$+$\gamma_3C_3(\Delta_3)$, where $\gamma_i$ denotes the ratio of QPs from each band and is roughly fixed as $\gamma_1:\gamma_2:\gamma_3=1:1:2$ based on the $C$($H$)/$T$ shown in Fig. 1(b). The data are well-fitted, as shown by the solid line giving the gap values $\sim$0.39 meV, 1.53 meV, and 1.64 meV for $\Delta_1$, $\Delta_2$, and $\Delta_3$, respectively. The smaller size of electronic $\Delta_1$ is qualitatively consistent with BQPI measurements \cite{BPQIarxiv}.

In summary, specific heat measurements considering field and field-angle dependence showed that FeSe consists of three gaps $\Delta_1$, $\Delta_2$, and $\Delta_3$. The smallest, $\Delta_1$, corresponding to the electron-type $\varepsilon$ band, has two vertical-line nodes (or gap minima) along the $k_z$ direction, and is isotropic out of the $ab$-plane. $\Delta_2$ is related to the electron-type $\delta$ band, and is isotropic in the $ab$-plane, but largely anisotropic out-of-plane. $\Delta_3$ residing on the hole-type $\alpha$ band shows small out-of-plane anisotropy with a strong Pauli-paramagnetic effect. These findings are helpful for understanding the exotic properties and unexpected high $T_{\rm c}$ in the FeSe system.

T. N. is supported by RIKEN Special Postdoctoral Researchers Program. The present work was supported by a Grant-in-Aid for Scientific Research on Innovative Areas ``J-Physics'' (15H05883) from MEXT, and KAKENHI (17K05553, 15K05158 and 17H01141) from JSPS.


\bibliography{ASHFeSereferences}

\end{document}